# Homonuclear J-Coupling Spectroscopy at Low Magnetic Fields using Spin-Lock Induced Crossing

Stephen J. DeVience*,[a] Mason Greer,[b] Soumyajit Mandal,[c] and Matthew S. Rosen[d]

[a] Dr. Stephen J. DeVience
Scalar Magnetics, LLC
Windsor Mill, MD 21244
E-mail: stephen@scalarmag.com

[b] Dr. Mason Greer
Department of Electrical, Computer, and Systems Engineering
Case Western Reserve University
Cleveland, OH 44106

[c] Prof. Soumyajit Mandal
Department of Electrical and Computer Engineering
University of Florida
Gainesville, FL 32611

[d] Prof. Matthew S Rosen
Athinoula A Martinos Center for Biomedical Engineering
Massachusetts General Hospital
Charlestown, MA 02129

Supporting information for this article is given via a link at the end of the document.

**Abstract:** Nuclear magnetic resonance (NMR) spectroscopy usually requires high magnetic fields to create spectral resolution among different proton species. At low fields, chemical shift dispersion is insufficient to separate the species, and the spectrum exhibits just a single line. In this work, we demonstrate that spectra can nevertheless be acquired at low field using a novel pulse sequence called spin-lock induced crossing (SLIC). This probes energy level crossings induced by a weak spin-locking pulse and produces a unique J-coupling spectrum for most organic molecules. Unlike other forms of low-field J-coupling spectroscopy, our technique does not require the presence of heteronuclei and can be used for most compounds in their native state. We performed SLIC spectroscopy on a number of small molecules at 276 kHz and 20.8 MHZ, and we show that SLIC spectra can be simulated in good agreement with measurements.

## Introduction

From its inception, NMR spectroscopy has experienced an uninterrupted trend toward increasing magnetic field strengths, which improves spectral resolution and sensitivity. Nevertheless, numerous applications exist where the use of low magnetic fields is desirable, such as in benchtop and educational instruments [1], portable operations for oil-field exploration [2], spectroscopy in the presence of ferromagnetic and paramagnetic substances [3], and optically-detected NMR with nitrogen vacancies as sensors [4]. Unfortunately, spectroscopy at low magnetic fields has classically been precluded by the nuances of MR physics. While spectral dispersion (chemical shift) is field-dependent, spin-spin couplings are not, and as the field is decreased these couplings come to dominate. At first, spectra start to become more complex as they stop following the simple rules of first-order perturbation theory predominant at high field. As the field is further decreased and spin-spin coupling becomes dominant, all spins become magnetically nearly-equivalent, and the spectrum of most molecules coalesces into a single spectral line providing no structural or identifying information (Fig. 1). This can occur even at moderate fields, 1 T and above, for many classes of molecules.

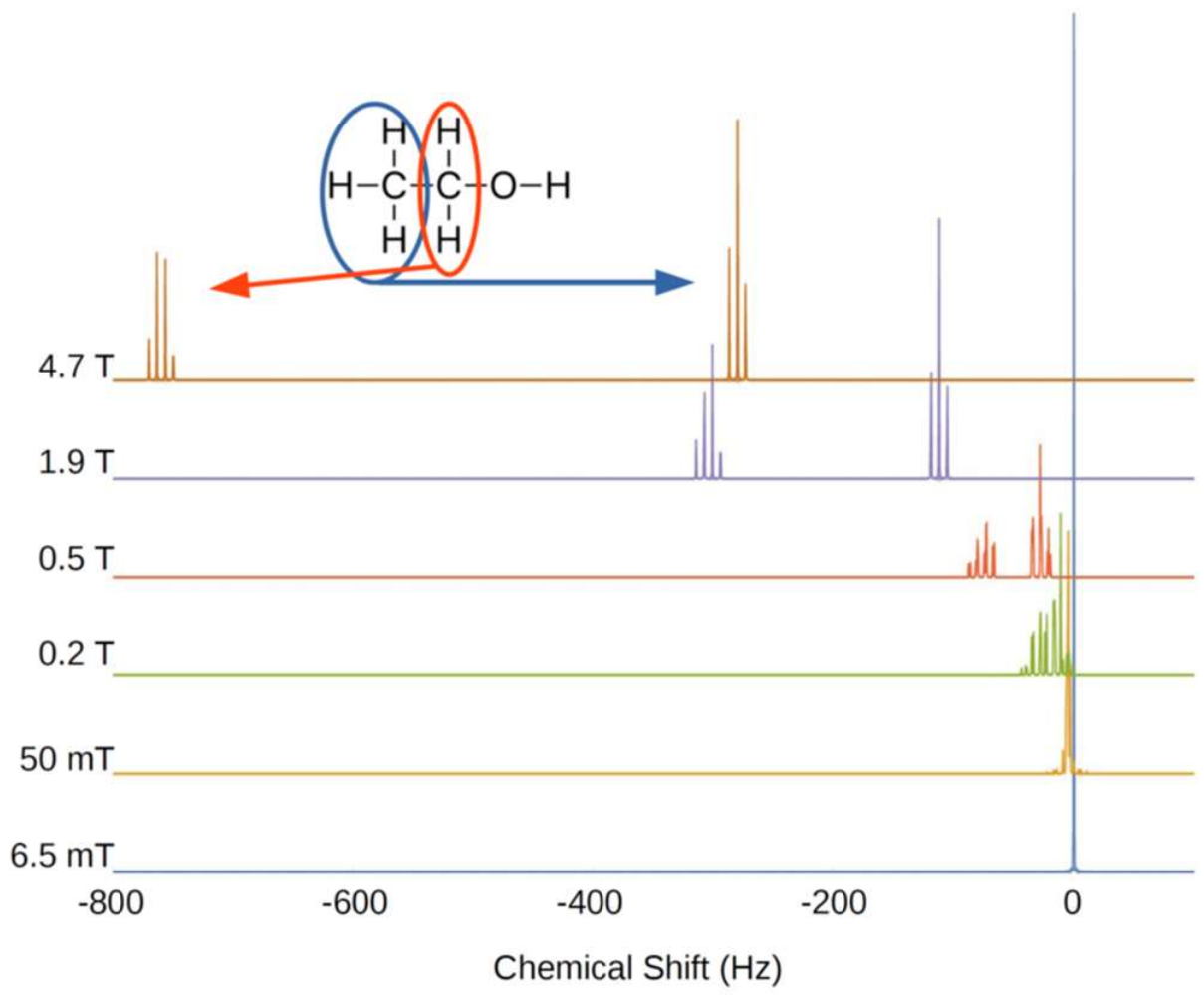

**Figure 1.** Simulated spectra of the methyl and methylene groups of ethanol as a function of magnetic field strength. Below 200 mT, the multiplets collapse into a single peak as J-coupling becomes stronger than chemical shifts.

A common work-around for this problem is to study substances containing a spin-1/2 heteronucleus, such as $^{13}C$, $^{15}N$, $^{19}F$, or $^{31}P$, which interacts with proton spins at low field to break magnetic near-equivalence and produce a complex J-coupling spectrum.[5] Instead of being separated by chemical shifts, the spectral lines reflect sums, differences, and multiples of the J-coupling strengths among spins, which are unique to each substance. However, the requirement of a coupled heteronucleus makes this technique impractical for most applications in organic chemistry, where hydrogen is often the only NMR-active nucleus. Even when a heteronucleus is present, they are usually spin>1/2, such as nitrogen and sulfur, and have relaxation times too short to create a J-coupling spectrum.

In this paper, we present an alternative approach to low-field spectroscopy that works for most homonuclear spin systems. Called SLIC spectroscopy, it is based on the spin-lock induced crossing method, which we previously utilized to manipulate singlet and triplet states in nearly-equivalent spin pairs, and to measure the spins' J-coupling and resonance frequency difference.[6] Here, we extend the work to arbitrary homonuclear spin systems having one or more resonance frequency difference, and we show that the pulse sequence produces a spectrum of dips at locations reflecting J-coupling strengths and molecular connectivity, which can be used to distinguish between compounds. These SLIC spectra can be simulated based on the chemical shifts and J-couplings known from high-field spectroscopy. We confirm these simulations with demonstrations of SLIC spectroscopy for numerous small molecules at 6.5 mT static field and for a series of chlorinated benzene compounds at 0.5 T.

## Theoretical Background

In high field NMR, the protons are normally under the condition $\delta\nu \gg J$, where $J$ is scalar coupling strength and $\delta\nu$ is the frequency difference between coupled spins. In this case, the spin system can be described by a product of Zeeman states such as $|\uparrow\uparrow\rangle$, $|\downarrow\uparrow\rangle$, etc. However, at low fields, where $J \gg \delta\nu$, the spin system must instead be described in terms of dressed states, i.e. superpositions of the Zeeman states. For the simplest system, a pair of coupled protons, these dressed states consist of three triplets and one singlet (Fig. 2a). They can be described by quantum numbers $|F, m_F\rangle$, where $F$ is the total spin quantum number and $m_F$ is the magnetic spin quantum number. In this notation, the singlet is $|0,0\rangle$ and the three triplets are $|1,1\rangle$, $|1,0\rangle$, and $|1,-1\rangle$. A conventional NMR sequence, such as a 90° pulse and FID, can only manipulate and detect transitions between $m_F$ states, and in this case would only detect transitions among the triplets.

The singlet state is thus unable to interact with the triplets in the conventional fashion. It is also separated from the triplets by a zero-field energy gap $J$. While a singlet-triplet coupling term does exist when $\delta\nu \neq 0$, it has no effect unless the singlet and triplet energy gap can be eliminated and the two states brought on resonance. This can be accomplished by applying continuous on-resonance spin–locking to the system. In the rotating frame, spin-locking creates rotated triplet states in the direction of $B_1$ and splits their energy levels proportionally to $B_1$ (quantified by the resulting nutation frequency, $\nu_n$). At the condition $\nu_n = J$, a triplet level is brought into resonance with the singlet, and coherent conversion between triplet magnetization and singlet order occurs.

To detect this singlet-triplet resonance, the SLIC spectroscopy sequence applies a $B_1$ pulse resonant with the conventional NMR peak for time $\tau_{SL}$. Multiple scans are performed, each time spin locking with a different $B_1$ and then acquiring an FID (Fig. 2b). When the singlet-triplet resonance condition occurs, some magnetization is converted to singlet order. This is detected as a decrease in the FID signal strength, or a decrease in the integral of the resulting spectral line. On a plot of integrated signal versus $\nu_n$, the intensity of the dip is proportional to $\sin(2\pi\tau_{SL}\delta\nu)$.

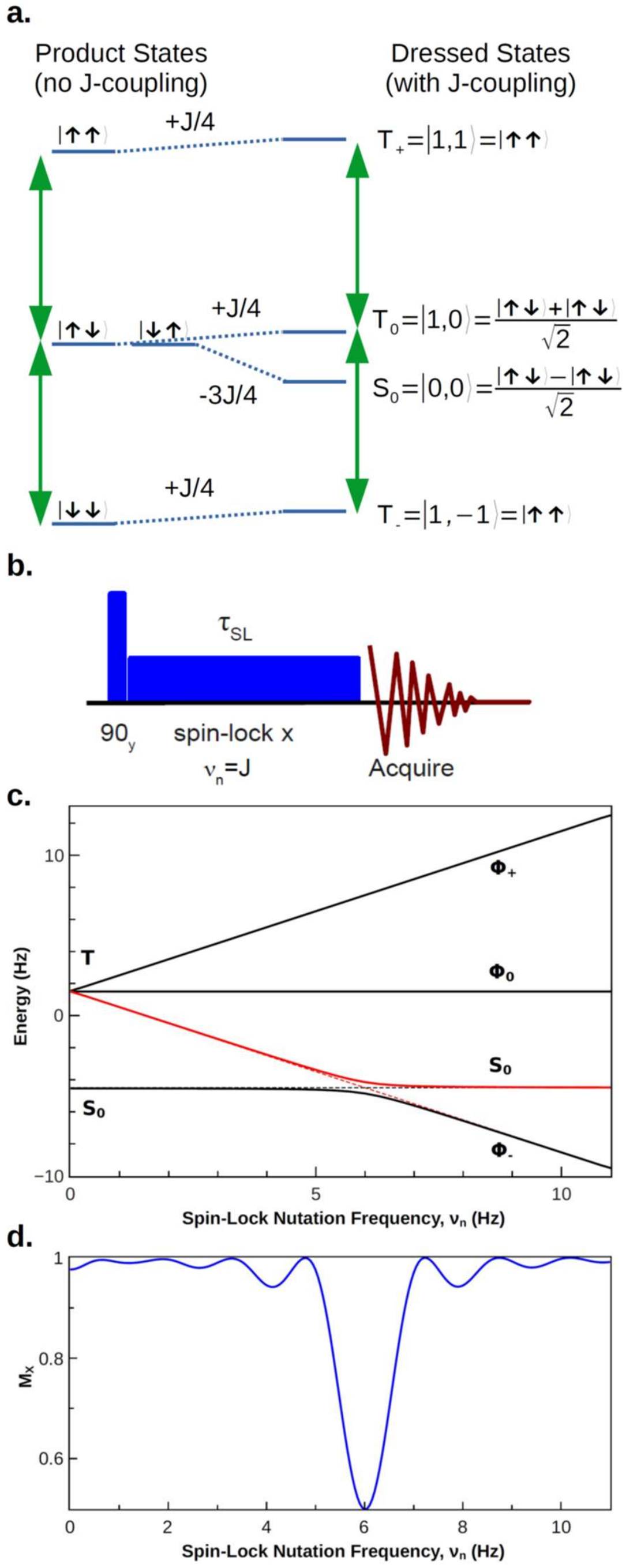

**Figure 2.** a) For two magnetically equivalent spins, J-coupling causes mixing of the energy levels and produces dressed states, consisting of three symmetric triplets and one antisymmetric singlet. The SLIC spectroscopy sequence (b) interrogates the energy levels (c) by perturbing the system with a weak spin-locking pulse on-resonance with the NMR spectral line. This induces a level anti-crossing where small chemical shift differences drive magnetization out of the x-axis, in this case from triplet states into the invisible singlet state. Multiple scans across a series of spin-lock nutation frequencies creates a spectrum (d) with a dip at the level anti-crossing, which in this simulation occurs at J=6 Hz.

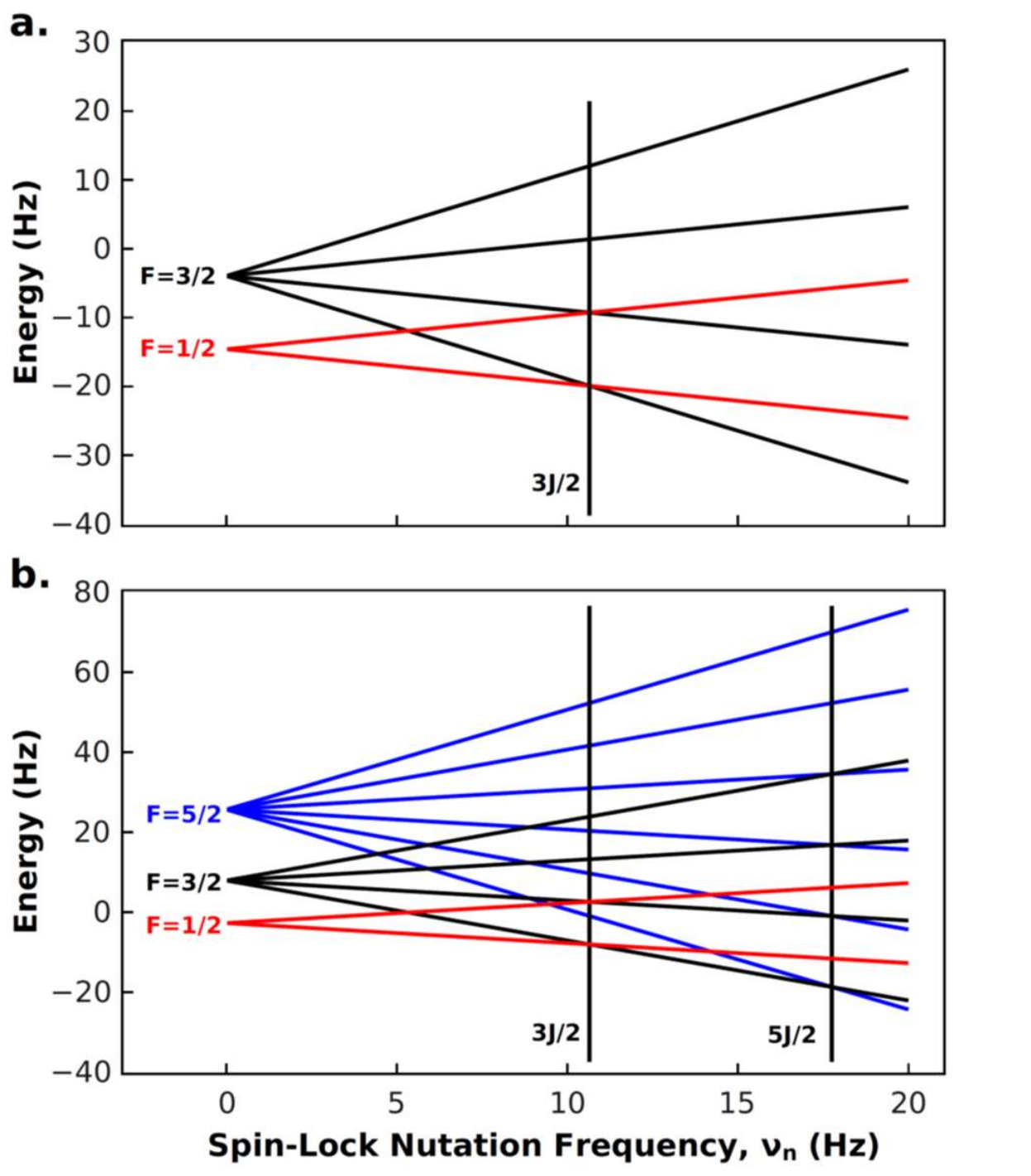

**Figure 3.** Energy levels during spin-locking of the hydrated ethanol system, which can be divided by symmetry properties into two groups (see section S1 of the supporting information). Anti-crossings occur at the locations indicated by vertical bars. Other crossings do not have interactions because they are not connected by the chemical shift Hamiltonian.

When more than two spins are nearly equivalent, dressed states of higher spin quantum number are formed, leading to a larger number of crossings and their associated dips. For example, figure 3 shows the predicted energy levels and crossings for the five-proton ethanol spin system (ignoring the alcohol proton). This $A_3B_2$ system occurs in ethyl acetate, 2-butanone, and hydrated ethanol undergoing fast exchange.

The level crossings can be determined analytically by diagonalizing the full Hamiltonian in the rotating frame and in the presence of the $B_1$ spin-locking pulse. One finds that the energy levels of the resulting eigenstates are determined by various sums and differences of $J$, each scaled by appropriate Clebsch-Gordan coefficients, and by a term $m_{FS}\nu_n = m_{FS}\gamma B_1/2\pi$ (here $m_{FS}$ is the magnetic spin quantum number of the rotated eigenstates). As shown in section S1 of the supplemental information, diagonalization leads to 32 eigenstates that can be subdivided into: (1) a group with maximum spin quantum number 5/2, having six $F = 5/2$ states, four $F = 3/2$ states, and two $F = 1/2$ states; (2) a group with maximum $F = 3/2$ having eight $F = 3/2$ states and four $F = 1/2$ states (some degenerate); (3) a group of eight states in which the methylene protons are in a singlet state. Under the influence of spin-locking, each of the first two groups experiences level anti-crossings among its states at which magnetization can be transferred. Magnetization transfer can only occur at crossings following selection rules $\Delta F = \pm 1$, and $\Delta m_{FS} = \pm 1$. For the $A_3B_2$ system, these crossings occur at $\nu_n = 3J/2$ and $5J/2$. The final group of spins does not play a role because there is no effective coupling with the methylene singlet states, and any crossings among the methyl proton states alone will not have an associated frequency difference to drive magnetization transfer ($\delta\nu_{AA} = 0$).

For more complex molecules, for example anhydrous ethanol where the alcohol must be taken into account, simulations were performed with a custom program written in MATLAB. The SLIC spectra were predicted based on literature values of J-couplings and chemical shifts acquired at high field.

## Results and Discussion

To confirm our predictions for the $A_3B_2$ system, we acquired SLIC spectra for ethyl acetate, 2-butanone, and hydrated ethanol at $B_0 = 276$ kHz. For all these molecules, $J_{AB} \approx 7.2$ Hz and $\delta\nu_{AB} \approx 0.7$ Hz. We also measured anhydrous ethanol, in which the alcohol proton does not experience exchange, and its coupling to the methylene protons must be taken into account.

The conventional NMR spectrum at 276 kHz exhibits a single peak with no features (see figure S1 of the supporting information). Figure 4 shows results for the measured SLIC spectra along with corresponding simulations using a spin-locking time of one second. The SLIC spectrum for ethyl acetate, 2-butanone, and hydrated ethanol all have two dips as predicted at $\nu_n = 3J/2$ and $5J/2$. From the dip locations, we measure $J_{AB} = 7.4$ Hz for ethyl acetate, $7.5$ Hz for 2-butanone, and $7.2$ Hz for hydrated ethanol. The SLIC spectra of these three compounds are similar because they have the same structural configuration ($A_3B_2$) among detectable spins. The isolated methyl groups of 2-butanone and ethyl acetate contribute to background signal but do not produce any dips, because they do not couple to the other groups in the molecule.

The dips are somewhat broader than the predicted spectra, and the intensity of the dips is about 0.1-0.2 units of $M_X$ lower than simulated, possibly because relaxation is not considered in the simulations. Broadening also appears to be related to the length of spin-locking versus the optimal time for coherent polarization transfer. The one second spin-locking time is 50% less than optimal for 2-butanone, but it is 10% longer than optimal for hydrated ethanol and 33% longer than optimal for ethyl acetate. In the simulations, this leads to wiggles in the latter two spectra, which cannot be resolved given the signal-to-noise, but may be contributing to the broadened lineshape. Overall, the linewidths were between 1.2 and 3.4 times broader than the predictions.

The ethanol spectrum is sensitive to the length of time the alcohol proton remains on the molecule. If the alcohol proton does not remain attached for sufficient time, for example due to very fast exchange with water, it does not effectively couple to the rest of the molecule via J-coupling, and $J_{AOH} = 0$ Hz. This results in the hydrated spectrum seen for 70% ethanol. For anhydrous ethanol, it remains attached for the entire measurement. This results in a molecule with a chain length one unit longer ($A_3B_2C$ configuration), leading to a very different SLIC spectrum. Anhydrous ethanol has four dips (Fig. 4d), and by matching to simulations we find $J_{AB} = 7.4$ Hz, as well as the coupling with the alcohol proton, $J_{AOH} = 5.4$ Hz. For intermediate exchange rates, the spectra must be calculated with more advanced methods, which will be discussed in a future paper.

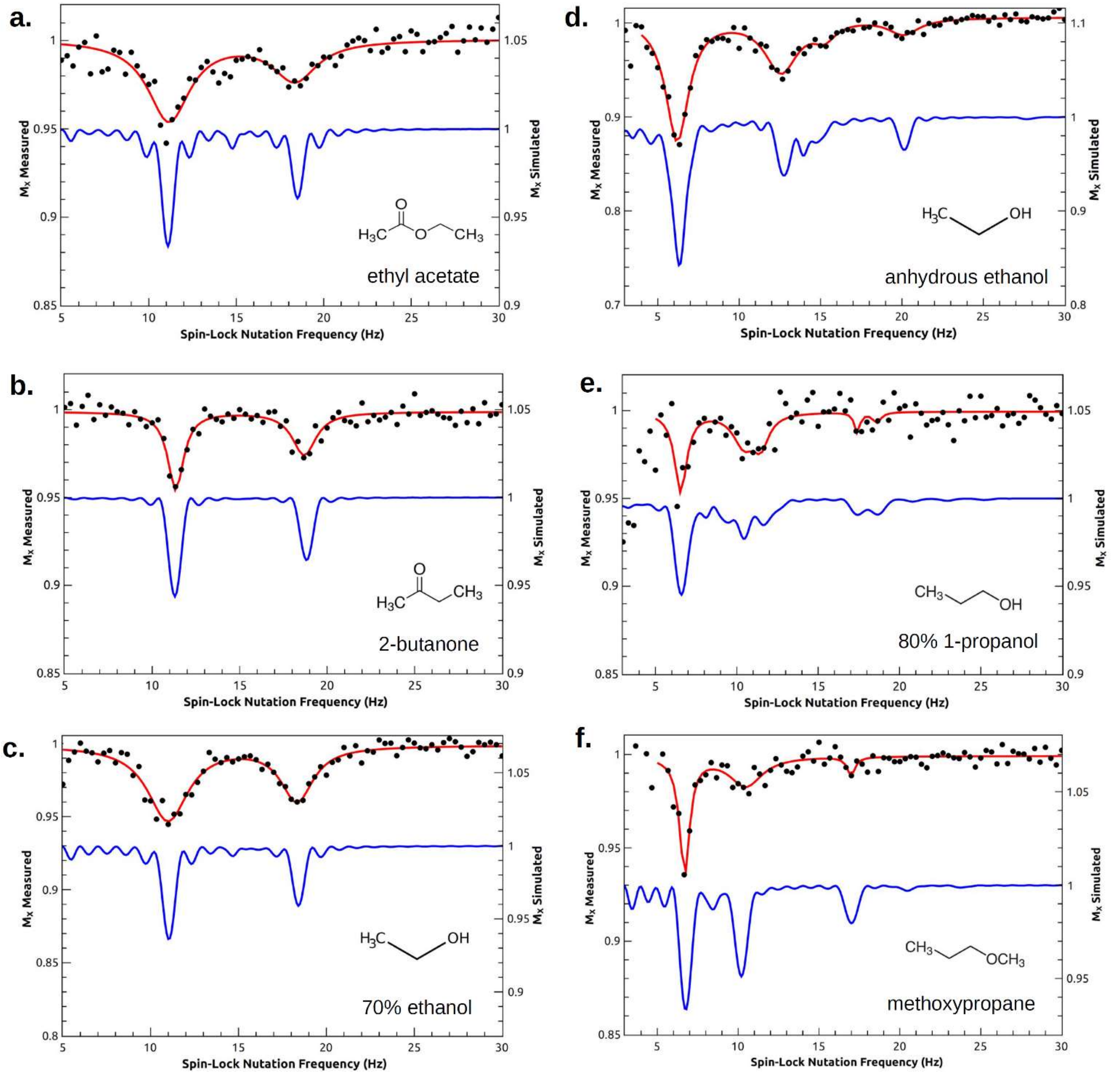

**Figure 4.** SLIC spectra for a number of compounds in the ethanol and 1-propanol families acquired at 6.5 mT ($^1$H frequency 276 kHz). Black points are measured data, red lines are best-fit curves using Lorentzian lineshapes for dips, and blue curves are simulated spectra. For the complex dips of anhydrous ethanol and 1-propanol, either one or two Lorentzian dips were used to achieve an approximate fit to the shape. Spectra are offset for comparison.

Adding a second spin in the third position to give the $A_3B_2C_2$ configuration, as in hydrated 1-propanol (Fig. 4e), produces a spectrum similar to anhydrous ethanol, with the strongest dip also near 6.5 Hz. The other dips are shifted downward compared to ethanol, meaning the extra C spin has the effect of compressing the spectrum toward lower frequencies. Methoxypropane has a similar structure (Fig. 4f), but it has a less complex spectrum than 1-propanol because of the smaller difference between $J_{AB}$ and $J_{BC}$ in the aliphatic chain. Literature values for $J_{AB}$ and $J_{BC}$ produced satisfactory results for both these compounds and were not adjusted. The simulations showed a strong sensitivity to $J_{AC}$, which was found to be about -0.2 Hz for 1-propanol and 0 Hz for methyoxypropane (see figure S2 of supporting information).

Figure 5 shows results for some other alcohols and ketones. The spectra for 1-butanol and 2-butanol show the continuing downward trend of the dip frequency and decrease in dip intensity as the chain length gets longer. Simulations of alkanes and other chains shows that this is a general limitation of the technique. As chains get longer, with numerous spins of similar chemical shift and J-couplings, the number of nearly degenerate energy levels increases exponentially and starts to create a continuum of levels. This is analogous to situations with Heisenberg chains as well as electronic energy levels in systems like long conjugated molecules.[7] Additionally, because of their shorter $T_1$ times, spin-locking was only applied for 750 ms for 1-butanol and 500 ms for 2-butanol, leading to broader dips. This is because the width and shape of the dip is determined by the Fourier transform of the SLIC pulse, with a longer pulse resulting in a narrower dip.

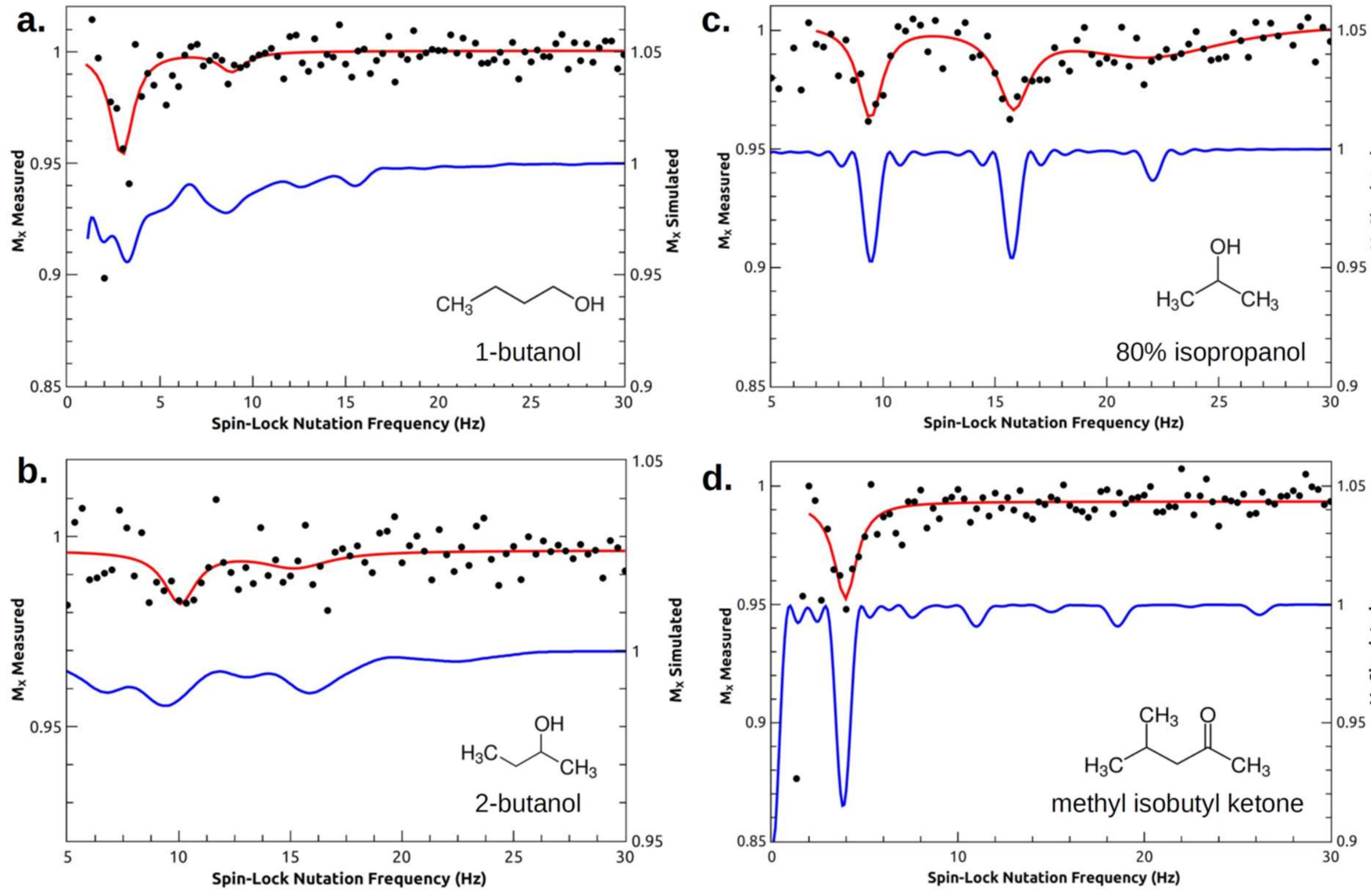

**Figure 5.** SLIC spectra for 1- and 2-butanol, hydrated isopropanol, and methyl isobutyl ketone acquired at 6.5 mT ([1]H frequency of 276 kHz). Black points are measured data, red lines are best-fit curves using Lorentzian lineshapes for dips, and blue curves are simulated spectra.

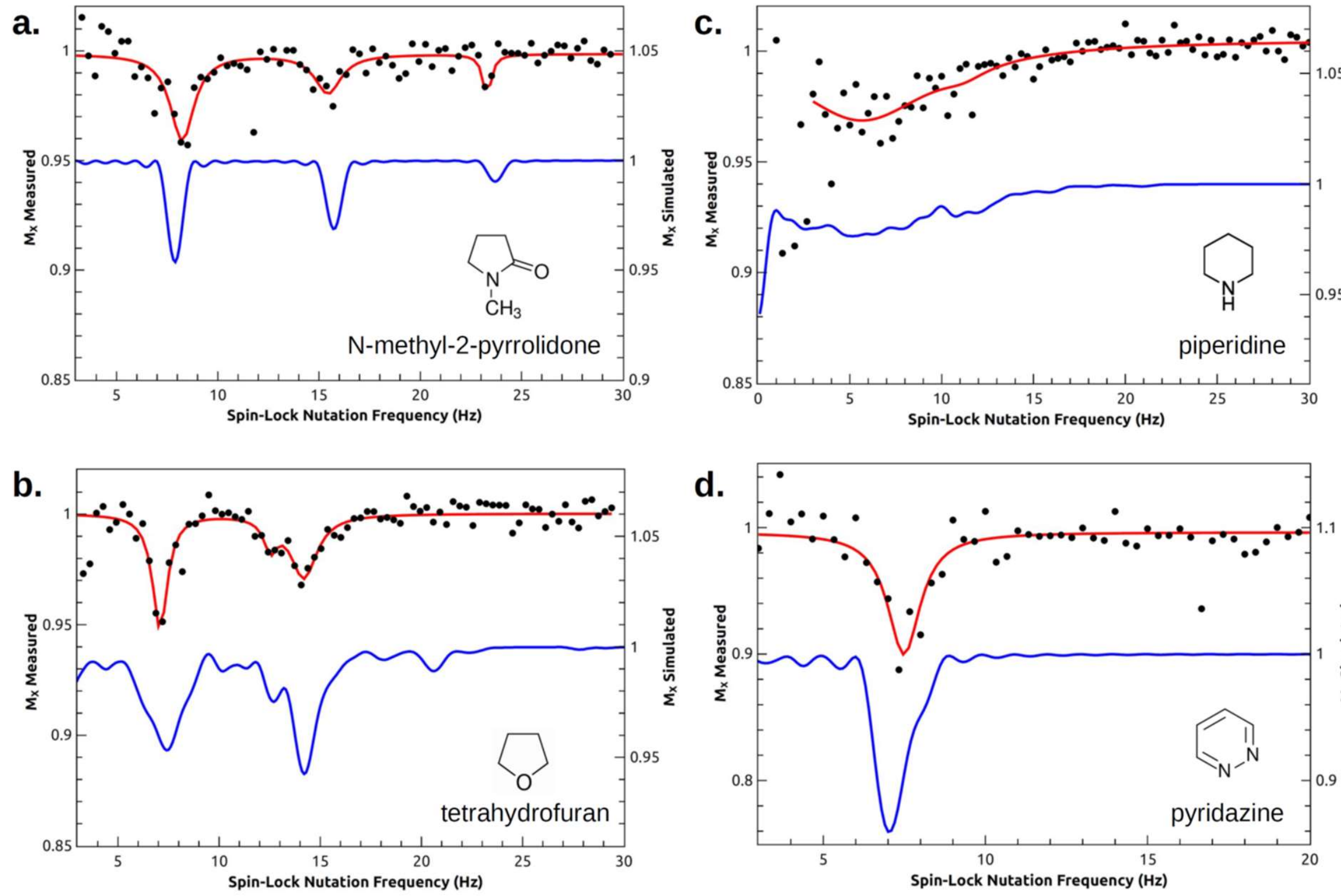

**Figure 6.** SLIC spectra for ringed compounds acquired at 6.5 mT ([1]H frequency 276 kHz). Black points are measured data, red lines are best-fit curves using Lorentzian lineshapes for dips, and blue curves are simulated spectra.

Hydrated isopropanol ($A_3BA'_3$) and methyl isobutyl ketone (($A_3BA'_3$)$C_2$) show some examples for symmetrically branched structures. Curiously, both measurements were missing a number of smaller dips at higher frequencies above the main dips, either due to insufficient SNR or some other unknown effect. For both these compounds, J-coupling needed to be adjusted upwards 5-9% to match the measured dip locations.

Figure 6 shows examples for ringed structures. N-methyl-2-pyrrolidone, tetrahydrofuran, and piperidine represent increasing lengths of proton chains from 3 through 5 pairs. N-methyl-2-pyrrolidone has a very similar spectrum to hydrated isopropanol, even though the spin systems are quite different ($A_2B_2C_2$ vs. $(A_3)_2B$). Similar to alkyl alcohols, as the chains get longer, the dips shift to lower frequencies and become increasingly complicated, and piperidine no longer shows any well-defined dips. Pyridazine, with only four protons, has a much simpler spectrum, but with the dip at about the same location as the first dip for its saturated analogue THF (configuration ABB'A vs. $A_2B_2B'_2A'_2$). Notably, although some of these molecules contain $^{14}N$, there was no effect of the quadrupolar spin coupling with the protons because the relaxation time of nitrogen is so short. Literature J-couplings produced good matches except for N-methyl-2-pyrrolidone, in which $J_{AB}$ needed to be adjusted from 7.2 to 7.8 Hz.

Another set of molecules investigated were isomers of dichloropropane and dichloropropene. 1,2-dichloropropane produced a very rich spectrum with five distinct dips, but required J couplings to be adjusted upwards 5-16% higher than literature values. The higher symmetry 1,3-dichloropropane ($A_2B_2A'_2$) produced three weaker dips at $J$, $2J$, and $3J$, giving $J_{AB} = 6.3$ Hz. This spectrum is similar to N-methyl-2-pyrrolidone, although in the latter the lower symmetry ($A_2B_2C_2$) leads to a different intensity pattern. Cis- and trans-1,3-dichloropropene also showed rich spectra in good agreement with literature values. At least three dips from each were not overlapping and could be used for determining relative concentration in a mixture of the two. Trans-1,3-dichloropropene had an extremely long $T_1$ of 9s, which led to well defined dips with high SNR. As with the nitrogen containing compounds, there was no noticeable effect of the quadrupolar chlorine nuclei.

Finally, SLIC spectra were acquired from four chlorinated benzene compounds at 20.8 MHz (~0.5 T): chlorobenzene, 1,2-dichlorobenzene, 1,3-dichlorobenzene, and 1,2,4-trichlorobenzene. For these, the chemical shift differences would be insufficient at 276 kHz to produce a reasonable dip contrast because chemical shifts would be on the order of 0.03 Hz. Even at 20.8 MHz the conventional spectrum consists of a single featureless line because chemical shifts are on the order of 2 Hz, smaller than the J-coupling. The resulting SLIC spectra are shown in Fig. 8. Spin-locking was only applied for 300 ms due to the relatively short $T_1$ of these compounds, as they are somewhat viscous at room temperature. The simulated spectra for chlorobenzene and 1,3-dichlorobenzene agree well with the measurements. For 1,2-dichlorobenzene, the high frequency dip is shifted about 2 Hz higher than predicted, and we were unable to account for this by adjusting J-couplings in the simulation. It might be due to miscalibration of the spin-locking power. For 1,2,4-trichlorobenzene, the dip at 13.5 Hz does not appear in the measured spectrum. For all four compounds, literature J-couplings produced main dip locations in reasonable agreement with measurements.

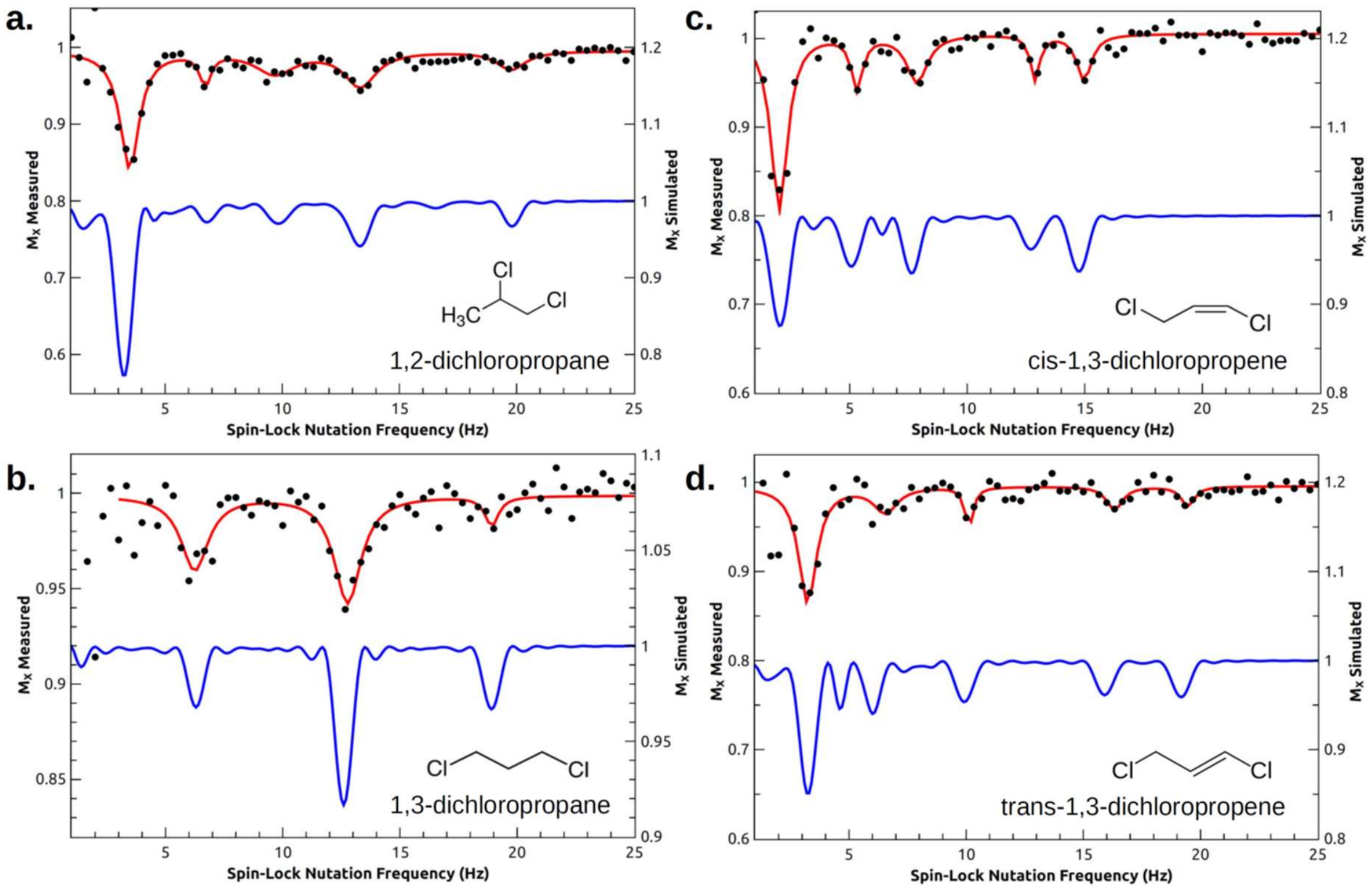

**Figure 7.** SLIC spectra for isomers of dichloropropane and dichloropropene acquired at 6.5 mT ($^1H$ frequency 276 kHz). Black points are measured data, red lines are best-fit curves using Lorentzian lineshapes for dips, and blue curves are simulated spectra.

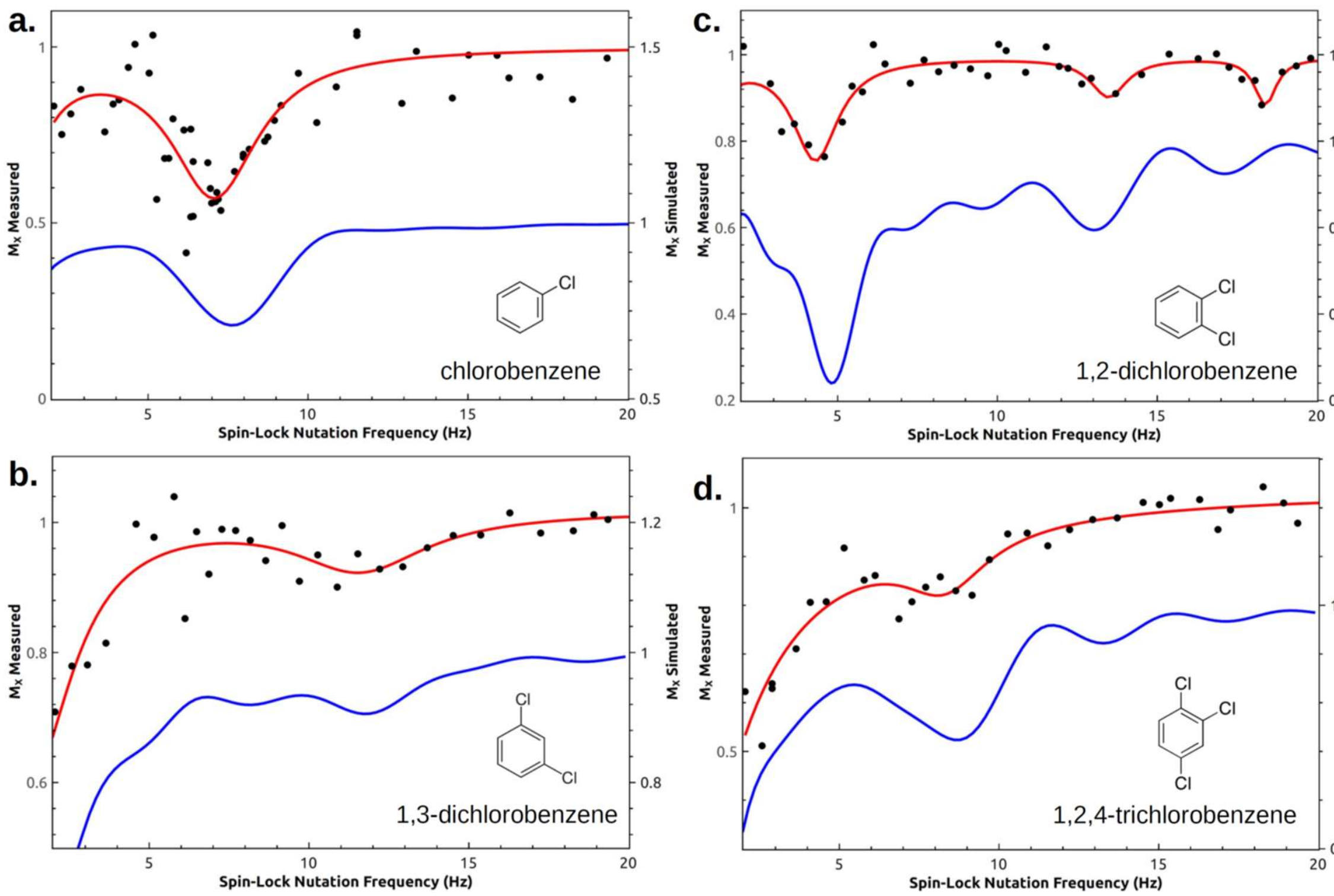

**Figure 8.** SLIC spectra for chlorinated benzenes acquired at 0.5 T ($^1$H frequency 20.8 MHz). Black points are measured data, red lines are best-fit curves using Lorentzian lineshapes for dips, and blue curves are simulated spectra.

Curiously, although both pyridazine and 1,2-dichlorobenzene share the ABB'A' spin configuration, they have quite different spectra. This might be because in pyridazine the $J_{BB'}$ coupling is significantly larger than $J_{AB}$ and $J_{A'B'}$ (8.2 Hz vs. 5 Hz), whereas these values are all similar in 1,2-dichlorobenzene (7.5 Hz and 8 Hz, respectively). This shows that for SLIC spectroscopy it is critical to perform a full simulation based on physical parameters, and that unlike conventional NMR spectroscopy, simple connectivity based rules are insufficient for spectral prediction.

Many of the SLIC measurements found J-coupling values higher than those from literature. This is likely because we used neat samples, whereas most literature spectra were acquired in deuterated chloroform. It is known that $J_{HH}$ couplings tend to increase along with solvent polarity, and neat alcohols are more polar than chloroform.[8] However, this cannot fully explain all the discrepancies, for example $J_{HH}$ of neat ethyl acetate has been measured to be 7.13 Hz at high field, yet we measure 7.4 Hz. It is possible our calibration procedure has some systematic error toward higher $B_1$. These differences are unlikely to be due to magnetic field dependence, as the difference in J-couplings between zero field and high field is expected to be at least two orders of magnitude smaller.[9]

One drawback of the technique is that SLIC-silent protons create a background signal on which the dips occur, meaning that when using SLIC in a mixture or in a protonated solvent, the dip intensity will be smaller than expected if the background protons are not taken into account. For example, acetone would create a large proton background and has no SLIC spectrum to even identify it as the solvent. This can be somewhat ameliorated by using deuterated solvents. It may also be possible to isolate the SLIC signal from the background via quantum filters.

Theoretically, a time dimension can also be acquired by taking spectra at a series of spin-locking times, which would produce oscillations proportional to chemical shift differences. An example is shown in figure S3 of the supplemental information. However, we found that at 276 kHz the signal decay due to $T_{1\rho}$ is on the same few-second timescale as these oscillations, making them difficult to measure. This is further complicated by the $B_1$ dependence of $T_{1\rho}$, because at smaller amplitudes CW spin-locking is less effective at overcoming decoherence.

## Conclusion

SLIC spectroscopy enables the identification and study of organic compounds via low-field NMR spectroscopy, even in the strong-coupling regime where the conventional NMR spectrum presents no identifying information. This may allow useful NMR spectra to be acquired with small inexpensive instruments using both conventional detection and new detection technology such as NV-diamond defects, which work better at low fields than at superconducting strengths. A number of questions also naturally arise from these results that require further investigation. For example, what are the lifetimes of the dressed states interrogated by SLIC spectroscopy, and because they incorporate spins from throughout the molecule, what can they tell us about molecular dynamics? How do different types of chemical exchange affect the SLIC spectra? How do heteronuclear spin couplings and related phenomena such as scalar relaxation affect the results?

## Experimental Section

SLIC spectra were simulated using custom code written in MATLAB. The algorithm diagonalizes the Hamiltonian in the presence of a $B_1$ field, propagates the time dependent Schrodinger equation, and measures the remaining x-axis magnetization, $M_x$. As a check, some simulations were also performed with the Spin Dynamica package in Mathematica[10]. However, Spin Dynamica was unable to handle more than a few spin system, and it was significantly slower because it is a general simulator for NMR dynamics and is not optimized to this particular problem. J-coupling and chemical shift parameters for the simulations were taken primarily from the SDBS website[11] along with other sources and are listed in the supporting information. As noted in the text, some J-coupling values were then adjusted to match simulations with measurements.

Samples were purchased from Sigma Aldrich (St. Louis, MO). For experiments at 276 kHz, samples were prepared neat in 10 mm diameter NMR tubes, unless otherwise noted. For measurements at 20.8 MHz, samples were prepared neat in 17 mm diameter by 60 mm long vials.

Spectra at 276 kHz were measured in a custom-built high-homogeneity electromagnet-based MRI scanner with a Tecmag Redstone console described previously.[12] For the presently described work, a solenoidal sample coil was used, designed to hold 10 mm NMR tubes, and a $B_0$ field-frequency lock was used to maintain the resonance frequency within ±0.25 Hz. The scanner was shimmed to achieve a linewidth of deionized water of better than 0.5 Hz. The extremely low power needed in these experiments was achieved by bypassing the transmit power amplifier, and RF pulses directly from the synthesizer were used, resulting in a 90˚ pulse length of 1 ms using about 4 μW. An active T/R switch was used to ensure the proper waveform of the low-power SLIC pulses. Typically, SLIC spectra were acquired with 8 averages using a four-step phase cycle. Spin-locking nutation frequency, $\nu_n$, was scanned with a 0.33 Hz step size. SLIC pulse length was 1 second unless otherwise noted.

Spectra at 20.8 MHz were acquired with a NUMAG 0.5 T MR magnet controlled by a Magritek Kea console. A custom built active T/R switch was used to switch between the high-power 90˚ pulse created with channel 1 via the power amplifier and the low-power SLIC pulses created directly from the channel 2 synthesizer. To correct for drift, the resonance frequency was adjusted at each acquisition to match the frequency of the previous FID. Nutation frequencies were chosen in a random order to avoid any additional bias due to drift. SNR was sufficient to acquire just a single measurement for each SLIC $B_1$ amplitude. SLIC pulse length was 300 ms.

In both systems, nutation frequency versus $B_1$ amplitude was calibrated by measuring the FID signal for a series of pulse lengths and then fitting the result with an exponentially decaying sinusoid function. After performing measurements at a number of RF amplitude values, a line was fit to the data to enable calculations for arbitrary amplitude. The relationship between nutation frequency and RF amplitude was linear for both systems. The calibration for the 276 kHz spectrometer was found to be stable over a time period of months and did not change significantly between samples.

For each spin-lock nutation frequency, the pulse sequence in Fig. 1b was played out, resulting in a FID readout. Each FID was converted to a spectrum via the fast Fourier transform, phase corrected at zero order, and integrated from -15 to 15 Hz. The integrals were divided by the maximal integrated signal from the whole set of spectra, and the result was plotted as a function of spin-lock nutation frequency to create a raw SLIC spectrum. The $T_{1\rho}$ background was then removed by dividing by a function

$$f(\nu_n) = A\left(1 - exp\left(-\frac{\nu_n}{B}\right)\right) + C\nu_n + D$$

where $\nu_n$ is the spin-lock nutation frequency, and *A, B, C,* and *D* are constants. Normally *A* was between 0 and 1, *B* was between 1 and 10, $C \approx 0$, and *D* was between 0 and 1.5. Finally, a sum of one or more Lorentzian dips was fit to the spectrum with least-squares fitting.

$T_1$ was acquired for each compound using an inversion recovery sequence to ensure the chosen spin-lock time did not exceed $T_1$ and to determine the delay time between SLIC acquisitions, which was set to 5 $T_1$. It may be possible to determine an optimal delay time for more time-efficient measurements in the future, similar to the Ernst angle. The measured $T_1$ values are listed in the supplemental information.

## Acknowledgements

This work was partially funded by NSF STTR contract 2014924. We are grateful for the assistance of Jarred Glickstein and David Ariando in setting up the 0.5 T experiment.

**Keywords:** NMR spectroscopy, spin-lock induced crossing, low-field NMR, J-coupling, dressed state

Supporting Information

# Homonuclear J-Coupling Spectroscopy at Low Magnetic Fields using Spin-Lock Induced Crossing

Stephen J. DeVience*, Mason Greer, Soumyajit Mandal, and Matthew S. Rosen

**Table of Contents**

**Section S1: Analytical calculation of ethanol energy levels during spin-locking**

For hydrated ethanol, the dressed states and their energy levels during spin-locking can be derived by hand.[1] The protons within the methylene group (labeled A) interact via scalar coupling with strength $J_{AA}$, while those within the methyl group (labeled B) interact with strength $J_{BB}$. These couplings are not measurable because the protons are magnetically equivalent, but the variables are retained for completeness. The methyl group protons are coupled to the methylene group protons by $J_{AB} \approx 7$ Hz. Because the methylene group protons are magnetically equivalent, they already exist in dressed states, specifically three degenerate triplets and one singlet:

$$
\begin{aligned}
|1,-1\rangle_A &= |\uparrow\uparrow\rangle & E &= \frac{1}{4}J_{AA} \\
|1,0\rangle_A &= \frac{|\uparrow\downarrow\rangle + |\downarrow\uparrow\rangle}{\sqrt{2}} & E &= \frac{1}{4}J_{AA} \\
|1,+1\rangle_A &= |\downarrow\downarrow\rangle & E &= \frac{1}{4}J_{AA} \\
|0,0\rangle_A &= \frac{|\uparrow\downarrow\rangle - |\downarrow\uparrow\rangle}{\sqrt{2}} & E &= -\frac{3}{4}J_{AA}.
\end{aligned}
\quad \text{(S1)}
$$

The methyl group protons are also magnetically equivalent and exist in eight dressed states. Four make up a spin-3/2 subspace, and the other four make up two spin-1/2 subspaces, which are degenerate but have different symmetries. States are designated by spin quantum numbers $F, m_F$. The spin-3/2 states are:

$$
\begin{aligned}
\left|\frac{3}{2},-\frac{3}{2}\right\rangle_B &= |\uparrow\uparrow\uparrow\rangle & E &= \frac{3}{4}J_{BB} \\
\left|\frac{3}{2},-\frac{1}{2}\right\rangle_B &= \frac{|\uparrow\uparrow\downarrow\rangle + |\uparrow\downarrow\uparrow\rangle + |\downarrow\uparrow\uparrow\rangle}{\sqrt{3}} & E &= \frac{3}{4}J_{BB} \\
\left|\frac{3}{2},+\frac{1}{2}\right\rangle_B &= \frac{|\downarrow\downarrow\uparrow\rangle + |\downarrow\uparrow\downarrow\rangle + |\uparrow\downarrow\downarrow\rangle}{\sqrt{3}} & E &= \frac{3}{4}J_{BB} \\
\left|\frac{3}{2},+\frac{3}{2}\right\rangle_B &= |\downarrow\downarrow\downarrow\rangle & E &= \frac{3}{4}J_{BB}.
\end{aligned}
\quad \text{(S2)}
$$

The spin-1/2 states are:

$$
\begin{aligned}
\left|\frac{1}{2},-\frac{1}{2}\right\rangle_B &= -\frac{2|\uparrow\uparrow\downarrow\rangle - |\uparrow\downarrow\uparrow\rangle - |\downarrow\uparrow\uparrow\rangle}{\sqrt{6}} & E &= -\frac{3}{4}J_{BB} \\
\left|\frac{1}{2},+\frac{1}{2}\right\rangle_B &= \frac{2|\downarrow\downarrow\uparrow\rangle - |\downarrow\uparrow\downarrow\rangle - |\uparrow\downarrow\downarrow\rangle}{\sqrt{6}} & E &= -\frac{3}{4}J_{BB} \\
\left|\frac{1}{2},-\frac{1}{2}\right\rangle_{BS} &= \frac{|\uparrow\downarrow\uparrow\rangle - |\downarrow\uparrow\uparrow\rangle}{\sqrt{2}} & E &= -\frac{3}{4}J_{BB} \\
\left|\frac{1}{2},+\frac{1}{2}\right\rangle_{BS} &= \frac{|\uparrow\downarrow\downarrow\rangle - |\downarrow\uparrow\downarrow\rangle}{\sqrt{2}}\rangle & E &= -\frac{3}{4}J_{BB}.
\end{aligned}
\quad \text{(S3)}
$$

The second pair of states are labeled S to designate that they are made up of singlet states of two spins paired with an up or down state of the third spin.

At 6.5 mT, the chemical shift between the methyl and methylene protons is much smaller than $J_{AB}$, and we will make the approximation that the chemical shift is zero. We combine the methyl and methylene product states into 32 new dressed states whose energies are now determine by all three terms $J_{AA}, J_{BB}, J_{AB}$. The internal energy states $J_{AA}$ and $J_{BB}$ will help determine which product states are grouped together. Mixing coefficients come from the Clebsch-Gorden coefficients.

The highest spin subspace is spin-5/2, all with energy $E = 3/2\, J_{AB} + 1/4\, J_{AA} + 3/4\, J_{BB}$.

$$\left|\frac{5}{2},-\frac{5}{2}\right\rangle = \left|\frac{3}{2},-\frac{3}{2}\right\rangle_B |1,-1\rangle_A$$
$$\left|\frac{5}{2},-\frac{3}{2}\right\rangle = \sqrt{\frac{2}{5}}\left|\frac{3}{2},-\frac{3}{2}\right\rangle_B |1,0\rangle_A + \sqrt{\frac{3}{5}}\left|\frac{3}{2},-\frac{1}{2}\right\rangle_B |1,-1\rangle_A$$
$$\left|\frac{5}{2},-\frac{1}{2}\right\rangle = \sqrt{\frac{1}{10}}\left|\frac{3}{2},-\frac{3}{2}\right\rangle_B |1,+1\rangle_A + \sqrt{\frac{3}{5}}\left|\frac{3}{2},-\frac{1}{2}\right\rangle_B |1,0\rangle_A + \sqrt{\frac{3}{10}}\left|\frac{3}{2},+\frac{1}{2}\right\rangle_B |1,-1\rangle_A$$
$$\left|\frac{5}{2},+\frac{1}{2}\right\rangle = \sqrt{\frac{1}{10}}\left|\frac{3}{2},+\frac{3}{2}\right\rangle_B |1,-1\rangle_A + \sqrt{\frac{3}{5}}\left|\frac{3}{2},+\frac{1}{2}\right\rangle_B |1,0\rangle_A + \sqrt{\frac{3}{10}}\left|\frac{3}{2},-\frac{1}{2}\right\rangle_B |1,+1\rangle_A$$
$$\left|\frac{5}{2},+\frac{3}{2}\right\rangle = \sqrt{\frac{2}{5}}\left|\frac{3}{2},+\frac{3}{2}\right\rangle_B |1,0\rangle_A + \sqrt{\frac{3}{5}}\left|\frac{3}{2},+\frac{1}{2}\right\rangle_B |1,+1\rangle_A$$
$$\left|\frac{5}{2},+\frac{5}{2}\right\rangle = \left|\frac{3}{2},+\frac{3}{2}\right\rangle_B |1,+1\rangle_A\,, \tag{S4}$$

The second subspace is spin-3/2, all with energy $E = -J_{AB} + 1/4\, J_{AA} + 3/4\, J_{BB}$:

$$\left|\frac{3}{2},-\frac{3}{2}\right\rangle = -\sqrt{\frac{3}{5}}\left|\frac{3}{2},-\frac{3}{2}\right\rangle_B |1,0\rangle_A + \sqrt{\frac{2}{5}}\left|\frac{3}{2},-\frac{1}{2}\right\rangle_B |1,-1\rangle_A$$
$$\left|\frac{3}{2},-\frac{1}{2}\right\rangle = -\sqrt{\frac{2}{5}}\left|\frac{3}{2},-\frac{3}{2}\right\rangle_B |1,+1\rangle_A - \sqrt{\frac{1}{15}}\left|\frac{3}{2},-\frac{1}{2}\right\rangle_B |1,0\rangle_A + \sqrt{\frac{8}{15}}\left|\frac{3}{2},+\frac{1}{2}\right\rangle_B |1,-1\rangle_A$$
$$\left|\frac{3}{2},+\frac{1}{2}\right\rangle = \sqrt{\frac{2}{5}}\left|\frac{3}{2},+\frac{3}{2}\right\rangle_B |1,-1\rangle_A + \sqrt{\frac{1}{15}}\left|\frac{3}{2},+\frac{1}{2}\right\rangle_B |1,0\rangle_A - \sqrt{\frac{8}{15}}\left|\frac{3}{2},-\frac{1}{2}\right\rangle_B |1,+1\rangle_A$$
$$\left|\frac{3}{2},+\frac{3}{2}\right\rangle = \sqrt{\frac{3}{5}}\left|\frac{3}{2},+\frac{3}{2}\right\rangle_B |1,0\rangle_A - \sqrt{\frac{2}{5}}\left|\frac{3}{2},+\frac{1}{2}\right\rangle_B |1,+1\rangle_A\,, \tag{S5}$$

And the third subspace is spin-1/2 with energy $E = -5/2\, J_{AB} + 1/4\, J_{AA} + 3/4\, J_{BB}$:

$$\left|\frac{1}{2},-\frac{1}{2}\right\rangle = \sqrt{\frac{1}{2}}\left|\frac{3}{2},-\frac{3}{2}\right\rangle_B |1,+1\rangle_A - \sqrt{\frac{1}{3}}\left|\frac{3}{2},-\frac{1}{2}\right\rangle_B |1,0\rangle_A + \sqrt{\frac{1}{6}}\left|\frac{3}{2},+\frac{1}{2}\right\rangle_B |1,-1\rangle_A$$
$$\left|\frac{1}{2},+\frac{1}{2}\right\rangle = \sqrt{\frac{1}{2}}\left|\frac{3}{2},+\frac{3}{2}\right\rangle_B |1,-1\rangle_A - \sqrt{\frac{1}{3}}\left|\frac{3}{2},+\frac{1}{2}\right\rangle_B |1,0\rangle_A + \sqrt{\frac{1}{6}}\left|\frac{3}{2},-\frac{1}{2}\right\rangle_B |1,+1\rangle_A\,, \tag{S6}$$

Note that all three of these make up a group containing the energy term $1/4\, J_{AA} + 3/4\, J_{BB}$.

A second group contains states with the energy term $\frac{1}{4}J_{AA} - 3/4\, J_{BB}$. There are two sets with spin-3/2 states and energy $E = \frac{1}{2}J_{AB} + \frac{1}{4}J_{AA} - 3/4\, J_{BB}$:

$$\begin{aligned}
\left|\frac{3}{2},-\frac{3}{2}\right\rangle_T &= \left|\frac{1}{2},-\frac{1}{2}\right\rangle_B |1,-1\rangle_A \\
\left|\frac{3}{2},-\frac{1}{2}\right\rangle_T &= \sqrt{\frac{1}{3}}\left|\frac{1}{2},+\frac{1}{2}\right\rangle_B |1,-1\rangle_A + \sqrt{\frac{2}{3}}\left|\frac{1}{2},-\frac{1}{2}\right\rangle_B |1,0\rangle_A \\
\left|\frac{3}{2},+\frac{1}{2}\right\rangle_T &= \sqrt{\frac{1}{3}}\left|\frac{1}{2},-\frac{1}{2}\right\rangle_B |1,+1\rangle_A + \sqrt{\frac{2}{3}}\left|\frac{1}{2},+\frac{1}{2}\right\rangle_B |1,0\rangle_A \\
\left|\frac{3}{2},+\frac{3}{2}\right\rangle_T &= \left|\frac{1}{2},+\frac{1}{2}\right\rangle_B |1,+1\rangle_A \\
\left|\frac{3}{2},-\frac{3}{2}\right\rangle_S &= \left|\frac{1}{2},-\frac{1}{2}\right\rangle_{BS} |1,-1\rangle_A \\
\left|\frac{3}{2},-\frac{1}{2}\right\rangle_S &= \sqrt{\frac{1}{3}}\left|\frac{1}{2},+\frac{1}{2}\right\rangle_{BS} |1,-1\rangle_A + \sqrt{\frac{2}{3}}\left|\frac{1}{2},-\frac{1}{2}\right\rangle_{BS} |1,0\rangle_A \\
\left|\frac{3}{2},+\frac{1}{2}\right\rangle_S &= \sqrt{\frac{1}{3}}\left|\frac{1}{2},-\frac{1}{2}\right\rangle_{BS} |1,+1\rangle_A + \sqrt{\frac{2}{3}}\left|\frac{1}{2},+\frac{1}{2}\right\rangle_{BS} |1,0\rangle_A \\
\left|\frac{3}{2},+\frac{3}{2}\right\rangle_S &= \left|\frac{1}{2},+\frac{1}{2}\right\rangle_{BS} |1,+1\rangle_A\,,
\end{aligned} \tag{S7}$$

and two sets with spin-1/2 states and energy $E = -J_{AB} + \frac{1}{4}J_{AA} - 3/4\, J_{BB}$:

$$\begin{aligned}
\left|\frac{1}{2},-\frac{1}{2}\right\rangle_T &= -\sqrt{\frac{2}{3}}\left|\frac{1}{2},+\frac{1}{2}\right\rangle_B |1,-1\rangle_A + \sqrt{\frac{1}{3}}\left|\frac{1}{2},-\frac{1}{2}\right\rangle_B |1,0\rangle_A \\
\left|\frac{1}{2},+\frac{1}{2}\right\rangle_T &= \sqrt{\frac{2}{3}}\left|\frac{1}{2},-\frac{1}{2}\right\rangle_B |1,+1\rangle_A - \sqrt{\frac{1}{3}}\left|\frac{1}{2},+\frac{1}{2}\right\rangle_B |1,0\rangle_A \\
\left|\frac{1}{2},-\frac{1}{2}\right\rangle_S &= -\sqrt{\frac{2}{3}}\left|\frac{1}{2},+\frac{1}{2}\right\rangle_{BS} |1,-1\rangle_A + \sqrt{\frac{1}{3}}\left|\frac{1}{2},-\frac{1}{2}\right\rangle_{BS} |1,0\rangle_A \\
\left|\frac{1}{2},+\frac{1}{2}\right\rangle_S &= \sqrt{\frac{2}{3}}\left|\frac{1}{2},-\frac{1}{2}\right\rangle_{BS} |1,+1\rangle_A - \sqrt{\frac{1}{3}}\left|\frac{1}{2},+\frac{1}{2}\right\rangle_{BS} |1,0\rangle_A
\end{aligned} \tag{S8}$$

The final 8 states consist of product states between the methyl group and the singlet state of the methylene group. Because $J_{AB}$ does not couple these states, they are not accessible to SLIC spectroscopy.

$$\begin{aligned}
&\left|\frac{3}{2},-\frac{3}{2}\right\rangle_B |0,0\rangle_A && E=\frac{3}{4}J_{BB}-\frac{3}{4}J_{AA}\\
&\left|\frac{3}{2},-\frac{1}{2}\right\rangle_B |0,0\rangle_A && E=\frac{3}{4}J_{BB}-\frac{3}{4}J_{AA}\\
&\left|\frac{3}{2},+\frac{1}{2}\right\rangle_B |0,0\rangle_A && E=\frac{3}{4}J_{BB}-\frac{3}{4}J_{AA}\\
&\left|\frac{3}{2},+\frac{3}{2}\right\rangle_B |0,0\rangle_A && E=\frac{3}{4}J_{BB}-\frac{3}{4}J_{AA}\\
&\left|\frac{1}{2},-\frac{1}{2}\right\rangle_B |0,0\rangle_A && E=-\frac{3}{4}J_{BB}-\frac{3}{4}J_{AA}\\
&\left|\frac{1}{2},+\frac{1}{2}\right\rangle_B |0,0\rangle_A && E=-\frac{3}{4}J_{BB}-\frac{3}{4}J_{AA}\\
&\left|\frac{1}{2},-\frac{1}{2}\right\rangle_{BS} |0,0\rangle_A && E=-\frac{3}{4}J_{BB}-\frac{3}{4}J_{AA}\\
&\left|\frac{1}{2},+\frac{1}{2}\right\rangle_{BS} |0,0\rangle_A && E=-\frac{3}{4}J_{BB}-\frac{3}{4}J_{AA}
\end{aligned} \tag{S9}$$

All of the previous states represent the dressed states at zero magnetic field strength with no spin-locking applied. Spin-locking further dresses the states with the RF Hamiltonian by rotating them into the x-axis (or whatever axis $B_1$ is along). In general, for spin-1/2, 3/2, and 5/2 groups the rotated states are:

$$\begin{aligned}
\left|\frac{1}{2},-\frac{1}{2}\right\rangle_{rot} &= -\frac{1}{\sqrt{2}}\left|\frac{1}{2},-\frac{1}{2}\right\rangle+\frac{1}{\sqrt{2}}\left|\frac{1}{2},+\frac{1}{2}\right\rangle\\
\left|\frac{1}{2},+\frac{1}{2}\right\rangle_{rot} &= \frac{1}{\sqrt{2}}\left|\frac{1}{2},-\frac{1}{2}\right\rangle+\frac{1}{\sqrt{2}}\left|\frac{1}{2},+\frac{1}{2}\right\rangle,
\end{aligned} \tag{S10}$$

$$\begin{aligned}
\left|\frac{3}{2},-\frac{3}{2}\right\rangle_{rot} &= -\frac{1}{2\sqrt{2}}\left|\frac{3}{2},-\frac{3}{2}\right\rangle+\frac{\sqrt{3}}{2\sqrt{2}}\left|\frac{3}{2},-\frac{1}{2}\right\rangle-\frac{\sqrt{3}}{2\sqrt{2}}\left|\frac{3}{2},+\frac{1}{2}\right\rangle+\frac{1}{2\sqrt{2}}\left|\frac{3}{2},+\frac{3}{2}\right\rangle\\
\left|\frac{3}{2},-\frac{1}{2}\right\rangle_{rot} &= \frac{\sqrt{3}}{2\sqrt{2}}\left|\frac{3}{2},-\frac{3}{2}\right\rangle-\frac{1}{2\sqrt{2}}\left|\frac{3}{2},-\frac{1}{2}\right\rangle-\frac{1}{2\sqrt{2}}\left|\frac{3}{2},+\frac{1}{2}\right\rangle+\frac{\sqrt{3}}{2\sqrt{2}}\left|\frac{3}{2},+\frac{3}{2}\right\rangle\\
\left|\frac{3}{2},+\frac{1}{2}\right\rangle_{rot} &= -\frac{\sqrt{3}}{2\sqrt{2}}\left|\frac{3}{2},-\frac{3}{2}\right\rangle-\frac{1}{2\sqrt{2}}\left|\frac{3}{2},-\frac{1}{2}\right\rangle+\frac{1}{2\sqrt{2}}\left|\frac{3}{2},+\frac{1}{2}\right\rangle+\frac{\sqrt{3}}{2\sqrt{2}}\left|\frac{3}{2},+\frac{3}{2}\right\rangle\\
\left|\frac{3}{2},+\frac{3}{2}\right\rangle_{rot} &= +\frac{1}{2\sqrt{2}}\left|\frac{3}{2},-\frac{3}{2}\right\rangle+\frac{\sqrt{3}}{2\sqrt{2}}\left|\frac{3}{2},-\frac{1}{2}\right\rangle+\frac{\sqrt{3}}{2\sqrt{2}}\left|\frac{3}{2},+\frac{1}{2}\right\rangle+\frac{1}{2\sqrt{2}}\left|\frac{3}{2},+\frac{3}{2}\right\rangle,
\end{aligned} \tag{S11}$$

$$\begin{aligned}
\left|\frac{5}{2},-\frac{5}{2}\right\rangle_{rot} &= -\frac{1}{4\sqrt{2}}\left|\frac{5}{2},-\frac{5}{2}\right\rangle + \frac{\sqrt{5}}{4\sqrt{2}}\left|\frac{5}{2},-\frac{3}{2}\right\rangle - \frac{\sqrt{10}}{4\sqrt{2}}\left|\frac{5}{2},-\frac{1}{2}\right\rangle \\
&+ \frac{\sqrt{10}}{4\sqrt{2}}\left|\frac{5}{2},+\frac{1}{2}\right\rangle - \frac{\sqrt{5}}{4\sqrt{2}}\left|\frac{5}{2},+\frac{3}{2}\right\rangle + \frac{1}{4\sqrt{2}}\left|\frac{5}{2},+\frac{5}{2}\right\rangle \\
\left|\frac{5}{2},-\frac{3}{2}\right\rangle_{rot} &= \frac{5}{4\sqrt{2}}\left|\frac{5}{2},-\frac{5}{2}\right\rangle - \frac{3}{4\sqrt{2}}\left|\frac{5}{2},-\frac{3}{2}\right\rangle + \frac{\sqrt{10}}{4\sqrt{2}}\left|\frac{5}{2},-\frac{1}{2}\right\rangle \\
&+ \frac{\sqrt{10}}{4\sqrt{2}}\left|\frac{5}{2},+\frac{1}{2}\right\rangle - \frac{3}{4\sqrt{2}}\left|\frac{5}{2},+\frac{3}{2}\right\rangle + \frac{5}{4\sqrt{2}}\left|\frac{5}{2},+\frac{5}{2}\right\rangle \\
\left|\frac{5}{2},-\frac{1}{2}\right\rangle_{rot} &= -\frac{\sqrt{10}}{4\sqrt{2}}\left|\frac{5}{2},-\frac{5}{2}\right\rangle + \frac{\sqrt{2}}{4\sqrt{2}}\left|\frac{5}{2},-\frac{3}{2}\right\rangle + \frac{2}{4\sqrt{2}}\left|\frac{5}{2},-\frac{1}{2}\right\rangle \\
&- \frac{2}{4\sqrt{2}}\left|\frac{5}{2},+\frac{1}{2}\right\rangle - \frac{\sqrt{2}}{4\sqrt{2}}\left|\frac{5}{2},+\frac{3}{2}\right\rangle + \frac{\sqrt{10}}{4\sqrt{2}}\left|\frac{5}{2},+\frac{5}{2}\right\rangle \\
\left|\frac{5}{2},+\frac{1}{2}\right\rangle_{rot} &= \frac{\sqrt{10}}{4\sqrt{2}}\left|\frac{5}{2},-\frac{5}{2}\right\rangle + \frac{\sqrt{2}}{4\sqrt{2}}\left|\frac{5}{2},-\frac{3}{2}\right\rangle - \frac{2}{4\sqrt{2}}\left|\frac{5}{2},-\frac{1}{2}\right\rangle \\
&- \frac{2}{4\sqrt{2}}\left|\frac{5}{2},+\frac{1}{2}\right\rangle + \frac{\sqrt{2}}{4\sqrt{2}}\left|\frac{5}{2},+\frac{3}{2}\right\rangle + \frac{\sqrt{10}}{4\sqrt{2}}\left|\frac{5}{2},+\frac{5}{2}\right\rangle \\
\left|\frac{5}{2},+\frac{3}{2}\right\rangle_{rot} &= -\frac{1}{4\sqrt{2}}\left|\frac{5}{2},-\frac{5}{2}\right\rangle - \frac{\sqrt{5}}{4\sqrt{2}}\left|\frac{5}{2},-\frac{3}{2}\right\rangle - \frac{\sqrt{10}}{4\sqrt{2}}\left|\frac{5}{2},-\frac{1}{2}\right\rangle \\
&+ \frac{\sqrt{10}}{4\sqrt{2}}\left|\frac{5}{2},+\frac{1}{2}\right\rangle + \frac{\sqrt{5}}{4\sqrt{2}}\left|\frac{5}{2},+\frac{3}{2}\right\rangle + \frac{1}{4\sqrt{2}}\left|\frac{5}{2},+\frac{5}{2}\right\rangle \\
\left|\frac{5}{2},+\frac{5}{2}\right\rangle_{rot} &= \frac{1}{4\sqrt{2}}\left|\frac{5}{2},-\frac{5}{2}\right\rangle + \frac{\sqrt{5}}{4\sqrt{2}}\left|\frac{5}{2},-\frac{3}{2}\right\rangle + \frac{\sqrt{10}}{4\sqrt{2}}\left|\frac{5}{2},-\frac{1}{2}\right\rangle \\
&+ \frac{\sqrt{10}}{4\sqrt{2}}\left|\frac{5}{2},+\frac{1}{2}\right\rangle + \frac{\sqrt{5}}{4\sqrt{2}}\left|\frac{5}{2},+\frac{3}{2}\right\rangle + \frac{1}{4\sqrt{2}}\left|\frac{5}{2},+\frac{5}{2}\right\rangle,
\end{aligned} \tag{S12}$$

The addition of spin-locking then perturbs these rotated states proportionally to their quantum number, i.e. by $\nu_n m_F$. When energy levels cross, the chemical shift Hamiltonian only produces level anti-crossings in certain situations. First, states from each of the two spin groups interact only with one another. So spin states from S4-S6 can interact, and those from S7 interact with S8. Second, there are selection rules that must be followed: $\Delta F_S = \pm 1, \Delta m_{FS} = \pm 1$, where these quantum numbers now refer to the rotated states. Taking this into account produces the levels and indicated crossings of Fig. 3 in the main text.

**Figure S1. Conventional NMR spectrum at 6.5 mT**

At 6.5 mT, the conventional NMR spectrum from a 90˚-FID sequence of neat ethyl acetate consists of a single narrow line. The two natural abundance $^{13}C$ satellites are also present, split by $J_{CH} = 128$ Hz. This spectrum was acquired with 16 averages and was processed with 1 Hz line broadening.

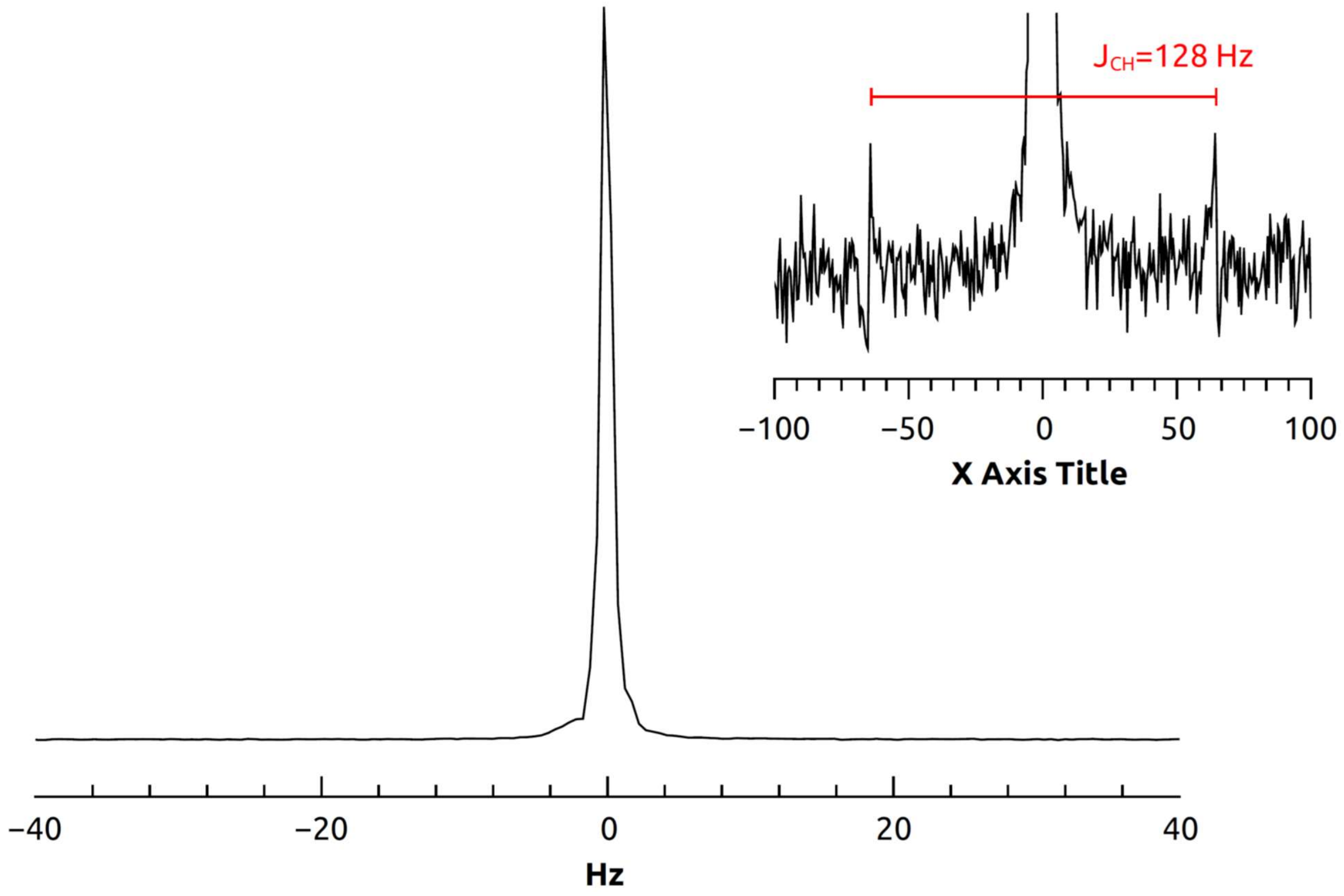

**Figure S2: Sensitivity to $^4J_{H-H}$**

SLIC spectra are sensitive to both the sign and magnitude of higher order J-couplings such as $^4J_{H-H}$. As an example, the figure below shows simulated spectra for hydrated 1-propanol with a range of $^4J_{H-H}$ values. In this case, the positions of the lowest and highest dips are the most sensitive.

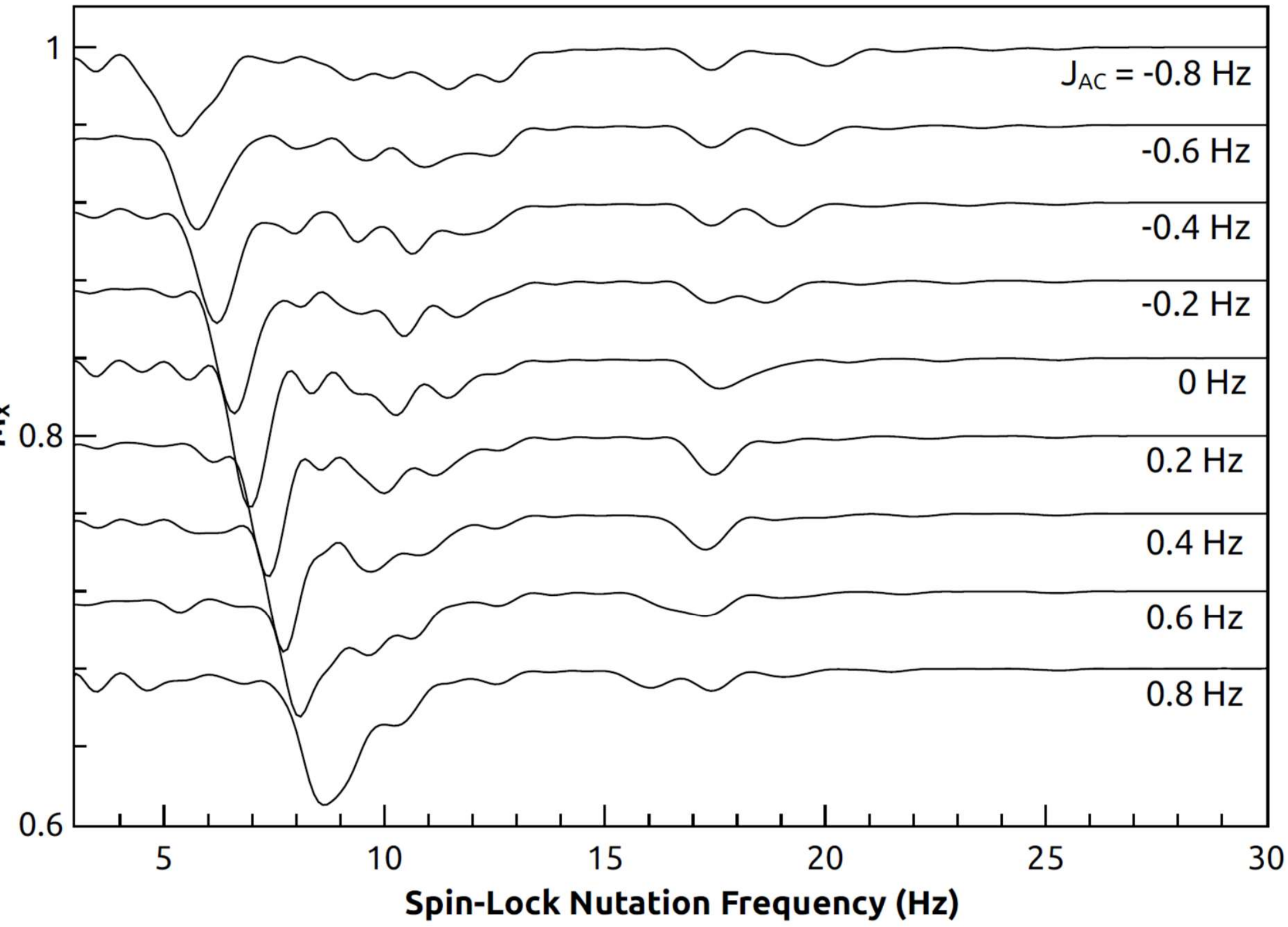

**Figure S3: Time domain measurements of ethyl acetate**

SLIC was performed on ethyl acetate with nutation frequency fixed to 11.3 Hz, 14 Hz, or 18.3 Hz and spin-locking time was varied from 0 to 2000 ms. Data from 11.3 Hz and 18.3 Hz were normalized by the values from 14 Hz, a frequency at which there is no dip. The measured curves (circles) show roughly the same oscillation patterns as simulations (line), but the contrast is lower. It could be that the nutation frequencies used were slightly off-resonant from the true dip resonance frequencies.

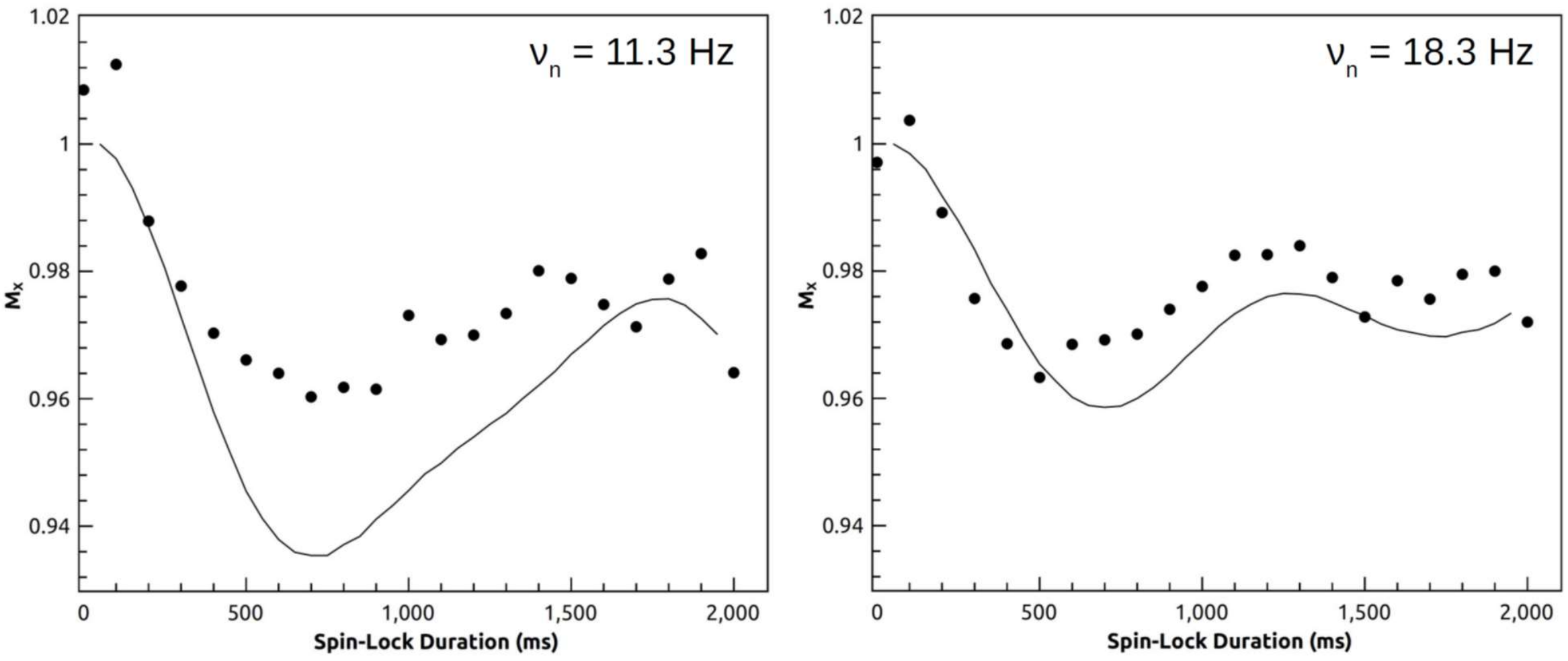

**Table 1. Simulation parameters and measured $T_1$**

Most chemical shifts and J-couplings were found in the Spectral Database System (SDBS).[1] Other sources are also indicated. For J-coupling, the literature values were used as a starting point, and the matched values were the final values determined to give good agreement with measured spectra. $T_1$ was measured at 6.5 mT with an inversion recovery sequence using delays of at least $5T_1$.

| Chemical | Chemical Shift | J Literature (Hz) | J Matched (Hz) | $T_1$ (s) |
|---|---|---|---|---|
| Ethyl acetate[1] | A 4.12<br>B 1.26<br>methyl 2.04 | $J_{AB}$=7.1 | $J_{AB}$=7.4 | 2.5 |
| 2-Butanone[1] | A 2.45<br>B 1.06<br>methyl 2.14 | $J_{AB}$=7.4 | $J_{AB}$=7.53 | 2.4 |
| Ethanol[1] | A 3.69<br>B 1.23<br>-OH 5.61 | $J_{AB}$=7.0<br>$J_{AOH}$=5.4 | $J_{AB}$=7.36<br>$J_{AOH}$=5.4 | 1.65 |
| 1-propanol[1] | A 3.44<br>B 1.48<br>C 0.88<br>-OH 2.58 | $J_{AB}$=6.6<br>$J_{BC}$=7.6<br>$J_{AOH}$=5.2 | $J_{AB}$=6.6<br>$J_{BC}$=7.6<br>$J_{AOH}$=0 (hydrated)<br>$J_{AC}$=-0.2 | 1.2* |
| Methoxypropane[1] | A 3.34<br>B 1.49<br>C 0.90<br>Methyl 3.34 | $J_{AB}$=6.7<br>$J_{BC}$=6.9 | $J_{AB}$=6.7<br>$J_{BC}$=6.9<br>$J_{AC}$=0 | 3.5 |
| 1-butanol[1-2] | A 3.63<br>B 1.53<br>C 1.39<br>D 0.94<br>-OH 2.24 | $J_{AB}$=6.36<br>$J_{AC}$=$J_{BD}$=-0.21<br>$J_{BC}$=6.12,8.66<br>$J_{CD}$=7.39<br>$J_{AOH}$=5.4 | $J_{AB}$=6.36<br>$J_{AC}$=$J_{BD}$=-0.21<br>$J_{BC}$=6.12,8.66<br>$J_{CD}$=7.39<br>$J_{AOH}$=5.4 | 0.76 |
| 2-butanol[1-2] | A 1.18<br>B 3.73<br>C 1.48<br>D 0.93<br>-OH 1.53 | $J_{AB}$=6.0<br>$J_{BC}$=6.0<br>$J_{AC}$=$J_{BD}$=-0.21<br>$J_{CD}$=7.4<br>$J_{BOH}$=5.4 | $J_{AB}$=6.6<br>$J_{BC}$=6.6<br>$J_{AC}$=$J_{BD}$=-0.21<br>$J_{CD}$=7.4<br>$J_{BOH}$=5.4 | 2.4 |
| Isopropanol[1] | A 3.87<br>B 1.10<br>-OH 2.57 | $J_{AB}$=6.0<br>$J_{AOH}$=4.4 | $J_{AB}$=6.3<br>$J_{AOH}$=0 (hydrated) | 1.2 |
| Methyl isobutyl ketone[1] | A 2.31<br>B 2.13<br>C 0.93<br>Methyl 2.12 | $J_{AB}$=7.2<br>$J_{BC}$=6.8 | $J_{AB}$=7.8<br>$J_{BC}$=7.4 | 1.9 |
| N-methyl-2-pyrrolidone[1] | A 3.41<br>B 2.04<br>C 2.36<br>Methyl 2.85 | $J_{AB}$=7.2<br>$J_{BC}$=8.0<br>$J_{CD}$=0.8 | $J_{AB}$=7.8<br>$J_{BC}$=8.0<br>$J_{CD}$=0.8 | 2.2 |
| Tetrahydrofuran[3] | A 3.74<br>B 1.85 | $J_{AA}=J_{A'A'}$=-8.13<br>$J_{BB}=J_{B'B'}$=-12.01<br>$J_{AB}=J_{A'B'}$=6.16,7.61<br>$J_{BB'}$=6.15,8.77<br>$J_{AB'}=J_{A'B}$=-0.49<br>$J_{AA'}$=-0.16 | $J_{AA}=J_{A'A'}$=-8.13<br>$J_{BB}=J_{B'B'}$=-12.01<br>$J_{AB}=J_{A'B'}$=6.16,7.61<br>$J_{BB'}$=6.15,8.77<br>$J_{AB'}=J_{A'B}$=-0.49<br>$J_{AA'}$=-0.16 | 3.3 |
| Piperidine[4] | A 2.79<br>B 1.52<br>C 1.48 | $J_{ax,eq}$=3.77<br>$J_{eq,eq}$=13.0<br>$J_{ax,ax}$=2.76 | $J_{ax,eq}$=3.77<br>$J_{eq,eq}$=2.76<br>$J_{ax,ax}$=13.0 | 2.2 |
| Pyridazine[1] | A 9.19<br>B 7.54 | $J_{AB}$=5.05<br>$J_{AB'}$=1.88<br>$J_{AA'}$=1.39<br>$J_{BB'}$=8.22 | $J_{AB}$=5.05<br>$J_{AB'}$=1.88<br>$J_{AA'}$=1.39<br>$J_{BB'}$=8.22 | 3.4 |
| 1,2-dichloropropane[1] | A 3.80<br>B 3.76<br>C 4.31<br>D 1.48 | $J_{AB}$=-11.07<br>$J_{AC}$=5.15<br>$J_{BC}$=5.90<br>$J_{CD}$=6.47 | $J_{AB}$=-11.07<br>$J_{AC}$=6.0<br>$J_{BC}$=6.6<br>$J_{CD}$=6.8 | 2.0 |
| 1,3-dichloropropane[5] | A 3.75<br>B 2.20 | | $J_{AB}$=6.3 | 2.0 |
| Cis-1,3-dichloropropene[6] | A 6.24<br>B 6.02<br>C 4.23 | $J_{AB}$=7.13<br>$J_{AC}$=-1.06<br>$J_{BC}$=7.38 | $J_{AB}$=7.13<br>$J_{AC}$=-1.06<br>$J_{BC}$=7.50 | 3.7 |
| Trans-1,3-dichloropropene[6] | A 6.32<br>B 6.05<br>C 4.01 | $J_{AB}$=13.17<br>$J_{AC}$=-1.27<br>$J_{BC}$=7.44 | $J_{AB}$=13.17<br>$J_{AC}$=-1.27<br>$J_{BC}$=7.44 | 9.4 |
| Chlorobenzene[1] | A 7.279<br>B 7.225<br>C 7.162 | $J_{AA'}$=2.27<br>$J_{AB}=J_{A'B'}$=8.05<br>$J_{AB'}=J_{A'B}$=0.44<br>$J_{AC}=J_{A'C}$ =1.17<br>$J_{BB'}$=1.70<br>$J_{BC}=J_{B'C}$ =7.46 | $J_{AA'}$=2.27<br>$J_{AB}=J_{A'B'}$=8.05<br>$J_{AB'}=J_{A'B}$=0.44<br>$J_{AC}=J_{A'C}$ =1.17<br>$J_{BB'}$=1.70<br>$J_{BC}=J_{B'C}$ =7.46 | |
| 1,3-dichlorobenzene[1] | A 7.270<br>B 7.104<br>C 7.051 | $J_{AB}=J_{AB'}$=1.97<br>$J_{AC}$=0.36<br>$J_{BB'}$=0.00<br>$J_{BC}=J_{B'C}$ =8.10 | $J_{AB}=J_{AB'}$=1.97<br>$J_{AC}$=0.36<br>$J_{BB'}$=0.00<br>$J_{BC}=J_{B'C}$ =8.10 | |
| 1,2-dichlorobenzene[1] | A 7.366<br>B 7.110 | $J_{AA'}$=0.32<br>$J_{AB}=J_{A'B'}$=8.04<br>$J_{AB'}=J_{A'B}$=1.54<br>$J_{BB'}$=7.45 | $J_{AA'}$=0.32<br>$J_{AB}=J_{A'B'}$=8.04<br>$J_{AB'}=J_{A'B}$=1.54<br>$J_{BB'}$=7.45 | |
| 1,2,4-trichlorobenzene[1] | A 7.36<br>B 7.12<br>C 7.27 | $J_{AB}$=2.20<br>$J_{AC}$=0.50<br>$J_{BC}$=8.20 | $J_{AB}$=2.20<br>$J_{AC}$=0.50<br>$J_{BC}$=8.20 | |

* For the mixture 80% 1-propanol and 20% water.